\documentclass[aps,pre,twocolumn,groupedaddress,superscriptaddress,showpacs]{revtex4-1} 
\usepackage{graphicx} 
\usepackage{color} 
\usepackage{amsmath} 
\usepackage{amsfonts} 
\usepackage{amssymb} 
\usepackage{dcolumn} 
\usepackage{hyperref} 
\hypersetup{colorlinks=true,urlcolor=blue,linkcolor=blue,citecolor=blue} 
\definecolor{OliveGreen}{RGB}{0,102,0} 
 
\begin{document} 
\title{Loop-erased random walk on a percolation cluster is compatible with Schramm-Loewner evolution 
} 
\author{E. Daryaei} 
\email{edaryayi@gmail.com} 
\affiliation{Department of Physics, Faculty of Basic Sciences, University of Neyshabur, P.O. Box 91136-899, Neyshabur, Iran} 
\begin{abstract} 
We study the scaling limit of planar loop-erased random walk (LERW) on the percolation cluster, with occupation probability $p\geq p_c$. We numerically demonstrate that the scaling limit of planar LERW$_p$ curves, for all $p>p_c$, can be described by Schramm-Loewner Evolution (SLE) with a single parameter $\kappa$ which is close to normal LERW in Euclidean lattice. However our results reveal that the LERW on critical incipient percolation clusters is compatible with SLE, but with another diffusivity coefficient $\kappa$. Several geometrical tests are applied to ascertain this. All calculations are consistent with $\mathrm{SLE}_\kappa$, where $\kappa=1.732\pm0.016$. This value of the diffusivity coefficient is outside of the well-known duality range $ 2 \leq \kappa \leq 8$. We also investigate how the winding angle of the LERW$_p$ crosses over from {\it Euclidean} to {\it fractal} geometry by gradually decreasing the value of the parameter $p$ from 1 to $p_c$. For finite systems, two crossover exponents and a scaling relation can be derived. We believe that this finding should, to some degree, help us to understand and predict the existence of conformal invariance in disordered and fractal landscapes. 
\end{abstract} 
\maketitle 
\section{Introduction} 
Anomalous diffusion in disordered media is a ubiquitous phenomenon in nature, ranging from 
physics and chemistry to biology and medicine \cite{Scher75, *Metzler01, *Ott90, *Min05, *Golding06, *Hall08}. The main feature of anomalous diffusion in disordered media is the fact that the mean square displacement of the diffusing species has a non-linear relationship with time \cite{Sahimi12, *Bouchaud90, *Havlin87}. Such disordered media is typically simulated through percolation systems, diffusion on percolation clusters have been studied in great detail \cite{Gefen83, Avraham00}. One could restrict the diffusion of a simple random walk (RW) to the incipient infinite cluster; It is known that, above criticality, $ p>p_c $, diffusion is anomalous over short distances and normal over long distances \cite{Avraham00}. However diffusion on critical incipient percolation clusters is anomalous in all length scales. 
On the other hand, one could erase the loops from trajectory of the RW chronologically this operation results in the loop erased random walk (LERW) \cite{Lawler80}. This model is equivalent to the uniform spanning trees \cite{Wilson96}, the q-state Potts model in the limit $q\rightarrow0$ \cite{LERW_exact}, and the avalanche frontier in Abelian sandpile model \cite{Majumdar92}. It is known that the fractal dimension of LERW in $D=2$ is $5/4$. Although scaling and universality class of LERW in integer lattice is known, the universality class of this model in fractal landscape and especially in critical percolation had not been hitherto studied. 
In addition to scale invariance and, consequently, fractal properties, it is well-known that 2D LERW is conformally invariant. This property causes that the measure of such 2D random curves to remain unchanged under transformations that preserve angles. Recent breakthrough of complex analysis has created a powerful tool for statistical characterization of conformal invariance of many discrete models in the scaling limit (see e.g. \cite{Cardy05} and references therein). In this new approach, now named Schramm-Loewner evolution (SLE), each random non-self-crossing curve, which possess conformal invariance and the domain Markov property, is mapped to a 1D Brownian motion on the real axis. Such Brownian motion has zero mean value and its variance grows linearly in time with a real positive coefficient $\kappa$ known as diffusivity \cite{Schramm00}. In this approach, all statistical properties of such 2D random curves (such as critical exponents and fractal dimension) can be obtained as functions of $\kappa$ \cite{Duplantier03, Cardy05}. Also due to a well-established relation between SLE and conformal field theory (CFT), a relationship between such 2D random curves and CFT models is possible \cite{Bauer06, Cardy05, central_charge}. So far, the SLE approach has been identified and studied theoretically and numerically in different statistical models such as in critical percolation \cite{Smirnov01}, self-avoiding walks \cite{Kennedy02}, Ising model \cite{Smirnov06}, spin glasses \cite{Amoruso06, Bernard07}, watershed \cite{Daryaei12}, and turbulence \cite{Bernard06}, as well as some other disordered models \cite{Schwarz09, Jacobsen09, Stevenson11, Chatelain10}. In particular, it has been proved that the scaling limit of LERW in a simply connected domain converges to SLE$_2$ \cite{Lawler04, Lawler11, Schramm00}. 
Establishing SLE for such systems has provided valuable information on the underlying symmetries and paved the way to some exact results \cite{Lawler01, Lawler11_1, Smirnov01, Smirnov06}. In fact, SLE is not a general property of non-self-crossing walks since many curves have been shown not to be SLE (for example see \cite{Risau-Gusman08, *Norrenbrock12}).

Recently, the scaling behavior of LERW on percolation cluster has been investigated \cite{Daryaei14}, As it has been rigorously proven recently, the scaling behavior of planar LERW$_p$, for all $p>p_c$, is the same as the LERW on Euclidean lattices \cite{Yadin11}. However, the LERW on critical percolation clusters scales with a new fractal dimension $d_f=1.217\pm0.002$ \cite{Daryaei14}. This fractal dimension clearly shows that this model is related to a family of curves appearing in different contexts such as, e.g., watershed of random landscapes \cite{Schrenk12, Cieplak94, Fehr11, *Fehr11b}, polymers in strongly disordered media \cite{Porto97}, invasion percolation \cite{Cieplak96}, bridge percolation \cite{Schrenk12}, and optimal path cracks \cite{Andrade09}.

By assuming translation, rotation, and scaling invariance for two-dimensional LERW$_p$ on percolation cluster, this question arises that whether SLE can be identified in such random curves. 
In the continuum limit of two-dimensional LERW$_p$ on percolation cluster one can check consistency with SLE process. The fractal dimension of the SLE$_{\kappa}$ curves is related to the diffusivity by the relation $d_f=min \lbrace 2,1+\frac{\kappa}{8}\rbrace$ \cite{Duplantier03, Beffara04, Cardy05}. If the LERW$_p$ is described by SLE process, then the diffusivity of them is given with the same relationship. Although it has been reported that the scaling limit of watersheds can be described by SLE \cite{Daryaei12}, it does not directly imply that the LERW on critical percolation is compatible with SLE because of the need for further conformal invariance and domain Markov property. 
 
In this paper we study the LERW on percolation cluster, with occupation probability above and equal to the critical value, $p\geq p_c$. 
Our results show that for all $p>p_c$, the scaling limit of obtained LERW$_p$ curves is close to exact results for LERW on Euclidean lattices first proposed by Schramm \cite{Schramm00}. To study the scaling limit of LERW$_p$ in two dimensions and compare it with SLE$_\kappa$, we carried out three different statistical evaluations, namely, the variance of the \emph{winding angle} (quantifying the angular distribution of the curves) \cite{Duplantier88, Wieland03}, the \emph{left-passage probability} \cite{Schramm00, Schramm01}, and the characterization of the driving function (\emph{direct SLE}) \cite{Bernard07}. We find that above the percolation threshold, i.e., $p>p_c$, all statistical evaluations are consist with $\kappa=2$. However, LERW on critical percolation are SLE curves of diffusivity $\kappa=1.732 \pm 0.016$. 
We simulate the LERW$_p$ on the percolation cluster as described in \cite{Daryaei14}. Then we show that the values of $\kappa$ independently obtained for each test are numerically consistent and in line with the fractal dimension of the LERW on critical percolation cluster. Hereafter, we discuss each analysis separately. 
\begin{figure}[h] 
\begin{center} 
\includegraphics[scale=0.42,angle=0]{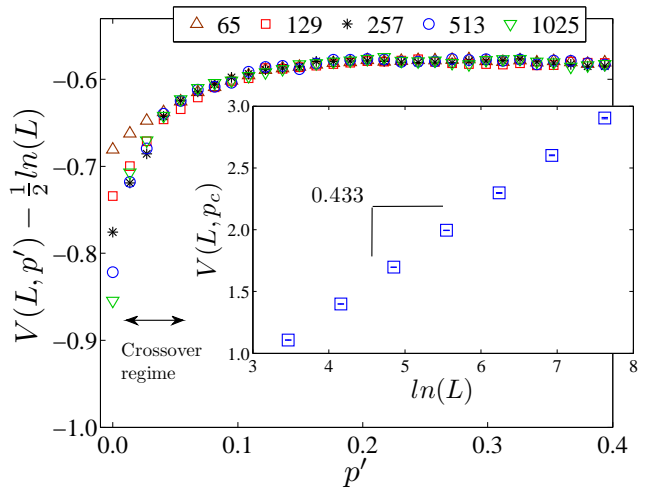} 
\end{center} 
\caption{(Color online) Deviation of winding angle variance of the LERW$_p$ on percolation cluster from normal LERW variance, i.e. $V(L,p^{\prime})-(\kappa/4) ln L$ as a function of $p^{\prime}$. \textbf{Inset}: Dependence of the winding angle variance on the lateral size of the lattice $L$ for LERW on critical percolation cluster (the statistical error bars are shown, but are quite shorter and appear as horizontal lines). The slope in the linear-log plot corresponds to $\kappa/4=0.433\pm 0.004$.} 
\label{fig::fig1} 
\end{figure} 
\section{Winding angle statistics} 
It is known that the winding angle distribution around a point for a 2D conformally invariant random curve can be related to Coulomb-gas parameter $g$ (which is directly related to the central charge $c=1-\frac{g(1-g)^2}{g}$ \cite{Francesco87}) and the system size $L$ \cite{Duplantier88}. The correspondence of the Coulomb-gas parameter $g$ to the relation for the winding angle variance can be extended to SLE \cite{Wieland03}. So we can test conformal invariance of the LERW$_p$ on percolation cluster and consistency with the SLE description by measuring the winding angle variance, as defined in \cite{Wieland03, Daryaei12}. The variance of the winding angle over all edges in the curve, $V(L,p)$ increases with the system size like $V(L,p)=b(p)+(\kappa/4) ln L$, where $b(p)$ is a constant which depends on the details of the definition \cite{Wieland03}. To measure the winding angle variance, we performed simulations at different lattice sizes; L = $2^{4+n}$ for $n=1,2,..6$. we generated $10^6$ LERW curves for small systems and more than $2\times 10^4$ for largest one. 
In the case of normal LERW ($p=1$), the winding angle variance of the curves logarithmically increases with system size as $V(L)\sim \frac{1}{2}ln L $, consistent with $\kappa=2$ of Euclidean LERW. By decreasing the occupation probability, $p$, the diffusivity coefficient of these random curves remains unchanged. At percolation threshold, these curves are smoother than normal LERW, and the winding angle variance increases logarithmically with the system size with different slope $V(L)\sim (\kappa/4) ln L$, with $\kappa\approx1.7$. Fig.~\ref{fig::fig1} shows dependence of the $V(L)-(1/2)ln L$ on $p^{\prime}$ for different system sizes, where $p^{\prime}$ is $p-p_c$. The overlap of the different curves confirms that the diffusion coefficient of the LERW$_p$ above $p_c$ is $2$. A small deviation is observed due to finite-size effects. There is a crossover between two different regimes near critical point $p\gtrsim p_c$ which can be observed in Fig.~\ref{fig::fig1}. 
At critical percolation, to obtain a more precise numerical estimation of $\kappa$, we increased the system size to $2^{11}$. The winding angle variance of LERW on critical percolation for different lattice size $L$ is shown in the inset of Fig.~\ref{fig::fig1}. We observe a slope of $0.433\pm0.004$ in a linear-log plot, which means that diﬀusivity is $\kappa=1.732 \pm 0.016$. This is in good agreement with the fractal dimension formula for SLE, i.e., $d_f=1+\kappa/8$ \cite{Beffara08}. 
\begin{figure} [!htb] 
\includegraphics[scale=0.53]{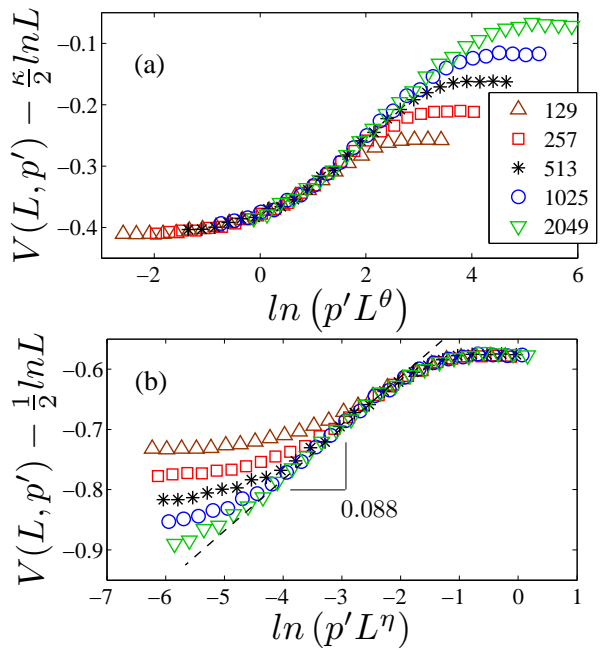} 
\caption{ (Color online) Crossover scaling and data collapse for LERW$_p$ in different system sizes. 
\textbf{(a)}: Deviation of winding angle variance of the LERW$_p$ on percolation cluster from variance of LERW on critical percolation cluster, i.e. $V(L,p^{\prime})-(0.433) ln L$ versus $ln\left(p^{\prime}L^{\theta}\right)$ for different system sizes. The scaling function given by equation~(\ref{eq::scaling1}) is applied, with $\theta=0.89\pm0.05$. 
\textbf{(b)}: Deviation of winding angle variance of the LERW$_p$ on percolation cluster from normal LERW variance, i.e. $V(L,p^{\prime})-\frac{1}{2} ln L$ versus $ln\left(p^{\prime}L^{\eta}\right)$, with $\eta=0.14\pm0.03$, for different system sizes. 
For each finite lattice size $L$, there are three regimes can be obtained with two crossover exponents i.e. $\theta$ and $\eta$. More precise estimate for $\beta$ can be obtained by the data collapsing different lattice sizes in intermediate regime which is $\beta=0.088\pm0.004$. 
All results have been averaged over $4\times10^4$ samples. 
\label{fig::fig2}} 
\end{figure} 
\section{Crossover scaling function} 
As shown in the Fig.~\ref{fig::fig1}, the winding angle variance of LERW$_p$ increases with increasing occupation probability. For large systems, the winding angle variance of LERW$_p$ grows with $p^{\prime}$, such that, $V(L,p^{\prime})\sim \beta ln\left( p^{\prime}\right)$, where $\beta\approx0.09$ is a novel coefficient, which we call variance-growth coefficient. There is a crossover behavior from {\it Euclidean} to {\it fractal} geometry \cite{Daryaei14}. Here, we try to investigate how the winding angle variance of the LERW$_p$ crosses over between these two universality classes by decreasing the value of the parameter $p$ from 1 to $p_c$. For the complete crossover scaling of the winding angle variance, $V$ can be considered as a logarithm of a homogeneous function on the relevant scaling fields, $V\left(bL,b^{y_p}p^{\prime}\right)=y_v ln b+V\left(L,p^{\prime}\right)$ where $b$ is a scaling parameter and $y_v$ and $y_p$ are relevant exponents for $V$ and $p$ scaling parameters respectively. One could restrict attention to the $p\rightarrow p_c$ regime, then for finite size of $L$, it is expected that $V$ increases with $\frac{\kappa}{4}$ slope, so in this regime $y_v=\frac{\kappa}{4}$. The next exponent can be found by trying to collapse the data (setting $b=L^{-1}$). The scaling {\it ansatz} for the winding angle variance is given by, 
\begin{equation} 
V\left(L,p^{\prime}\right)=ln \left( L^{\frac{\kappa}{4}} \mathcal{G}\left[p^{\prime}L^{\theta}\right]\right) \ \ , \label{eq::scaling1} 
\end{equation} 
\noindent where $\mathcal{G}\left[u\right]$ is a scaling function, such that, $\mathcal{G}\left[u\right]\sim u^{\beta}$ for small values of $u$, and is nonzero at $u\rightarrow0$. The exponent $\theta=y_p$ is the crossover exponent in the $p\rightarrow p_c$ regime. 
Fig.~\ref{fig::fig2}(a) shows crossover scaling for different lattice sizes, close to the critical point. 
As it shown, we have a good data collapse for small values of $u$ with $\theta=0.89\pm0.05$. For each finite lattice size $L$, there is a crossover point such as $p_{\times_1}^{\prime}$ scales like $L^{-\theta}$, which for $u \ll1$ we have a saturation regime, and for $u\gg1$ results are consistent with $\beta ln(u)$ for all lattice size $L$. However for large values of $p^{\prime}L^{\theta}$, we don't observe data collapse and the winding angle variance behaves as $\frac{(2-\kappa)}{4}ln (L)$. 
On the other hand, one could look large values of $p$, it is expected that the winding angle variance behaves like Euclidean geometry, so $y_v=\frac{1}{2}$ in this regime. If we follow the same strategy as above, we could find another scaling function; 
\begin{equation}\label{eq::scaling} 
V\left(L,p^{\prime}\right)=ln \left( L^{\frac{1}{2}}\mathcal{F}\left[p^{\prime}L^\eta\right]\right), 
\end{equation} 
where the scaling function $\mathcal{F}[x]$ have a saturation regime for large values of $x$, and the exponent $\eta=y_p$ is the corresponding crossover exponent in this regime. In fact, we could find another crossover point, $p_{\times_2}^{\prime}$ scaling with $L^{-\eta}$ which the winding angle variance behaves like $ln(\mathcal{F}[x])\sim\beta ln(x)$ for $x\ll 1$, and is a $const$ value for $x\gg 1$. Fig.~\ref{fig::fig2}(b) shows the scaling behaviors for different lattice size of $L$. As it shown, we have a good data collapse with $\eta=0.14\pm 0.03$, this clearly shows that the argument of $p^{\prime}L^{\eta}$ in crossover point should be independent of lattice size, so the crossing probability $p_{\times_2}^{\prime}$ scales like $L^{-\eta}$ with system size. 
The overlap of the different curves confirms that the diffusion coefficient of the LERW$_p$ above $p_c$ is $2$. 
Three different regimes, as shown in Fig.~\ref{fig::fig2}, are clearly identified; for $p^{\prime}<p_{\times_1}^{\prime}$ the winding angle variance behaves $V\sim (0.433) ln (L)$, for $p_{\times_1}^{\prime}<p^{\prime}<p_{\times_2}^{\prime}$, $S$ has a logarithmic behavior as $\beta ln (p^{\prime})$, and finally for $p_{\times_2}^{\prime}<p^{\prime}$, it behaves with Euclidean exponent, i.e. $\sim \frac{1}{2} ln (L)$. Therefore, the following relation can be derived, 
\begin{equation}\label{eq::scaling.relation} 
\beta\left(\theta-\eta\right)=\frac{1}{4}\left(2-\kappa \right), 
\end{equation} 
which is in good agreement with our obtained numerical values for the exponents. Interestingly, by considering $\kappa=8(d_f-1)$ for SLE curves, this relation is consistent with reported scaling relation for mean total length of LERW$_p$ \cite{Daryaei14}. 
\begin{figure}[!htb] 
\includegraphics[scale=0.41]{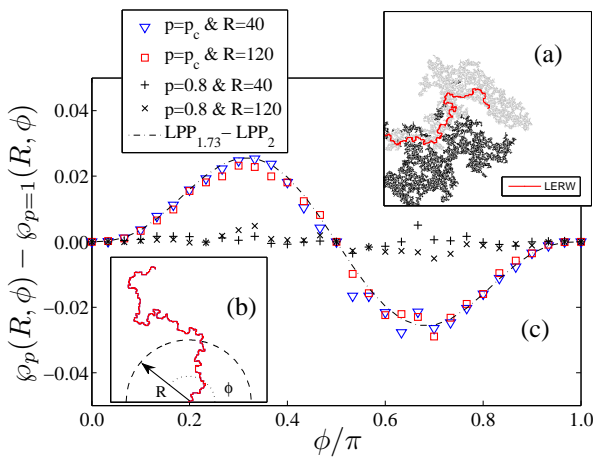} 
\caption{\label{left}(Color online) Left-passage probability for LERW$_p$ on percolation clusters as function of polar angle. \textbf{(a)} LERW on critical percolation cluster (shown in black color) on $513\times513$ lattice (the visited sites are shown in gray). \textbf{(b)} A schematic representation of the left-passage definition (details in the text) on obtained curve after rotation and translation. \textbf{(c)} $\wp_{p}(\phi,R)-\wp_{p=1}(\phi,R)$ for the LERW in the upper-half plane where $\wp_{p}(\phi,R)$ is the probability for the LERW on percolation cluster to pass to the left of a point with polar coordinates $(R,\phi)$. At $p=p_c$ the results are in good agreement with $LPP_{1.73}-LPP_2$ where $LPP_{1.73}(\phi)$ is the left-passage probability for SLE$_{1.73}$ given by Schramm's formula, Eq.~(\ref{left_formula}). The magnitude of statistical errors (not shown) is consistent with the apparent fluctuations of the data lines. The results are averages over $5\times10^4$ curves on a square lattice with $L=513$.} 
\end{figure} 
\section{Left passage probability} 
By considering scale invariance of SLE$_\kappa$ curves in the upper half-plane $\mathbb{H}$, one can determine the probability that a point $R e^{i\phi}$ is at the right side of the curve (see Fig.~\ref{left}(b)). This probability only depends on $\phi$ and is given by Schramm's formula \cite{Schramm00}, 
\begin{equation} 
LPP_{\kappa}(\phi)=\frac{1}{2}+\frac{\Gamma\left(\frac{4}{\kappa}\right)}{\sqrt{\pi} 
\Gamma\left(\frac{8-\kappa}{2\kappa}\right)}\cot(\phi)_2F_1 
\left(\frac{1}{2};\frac{4}{\kappa},\frac{3}{2};-\cot^{2}(\phi)\right)\,,\label{left_formula} 
\end{equation} 
where $_2F_1$ is the hypergeometric function and $\Gamma$ is the Gamma function. This formula is valid only for the upper half plane domain that SLE curve starts from the origin and goes to infinity, i.e., {\it chordal} SLE. In order to simulate LERW on upper half plane, it can be obtained by some rotations and translations of the curves described above on the whole plane to the upper-half plane for details see \footnote{Simple transformations of rotations and translations were done as follow: we clockwisely rotate the curve around the ending point on boundary by $-\pi/2$,$\pi/2$,$\pi$, and $0$ for ending point on right, left, top, and bottom boundary respectively, then the rotated curves were translated to the origin.}. In Fig.\ \ref{left}(a) and (b) a random LERW curve before and after transformations is shown. It is important to notice that these curves are not {\it chordal} completely due to restriction of going to middle of the lattice, However they behave like {\it chordal} near their starting points \footnote{The transformed curve is {\it dipolar} which starts from origin and goes to a random point on $y=L/2$, so our measurement should be done only for small radius for {\it dipolar} curves as discussed in \cite{Bernard07}. Our results are independent of $R$ values for $R<L/4$}. 
We measure the left passage probability, $\wp_{p}(\phi,R)$ for the LERW$_p$ curves in three different occupation probability, $p=1$, $0.8$, and $p_c$. We can reduce both the finite-size effects and other effects which are related to not being in chordal type by comparing with normal LERW on same lattice size, i.e., $\Delta\wp(\phi,R,p)=\wp_{p}(\phi,R)-\wp_{p=1}(\phi,R)$. Our results for $p=0.8$ (as a sample in {\it Euclidean} regime) and $p=p_c$ are shown in Fig.\ref{left}. The comparing left passage probability of LERW curves on critical percolation, i.e., $\Delta\wp(\phi,R,p_c)$, is in good agreement with $LPP_{1.73}(\phi)-LPP_2(\phi)$. As it is shown in Fig.\ \ref{left} this quantity is independent of $R$ values and consequently our results for LERW on critical percolation are consistent with the SLE$_{1.73}$. 
\section{Direct SLE test} 
\begin{figure}[h] 
\includegraphics[scale=0.47]{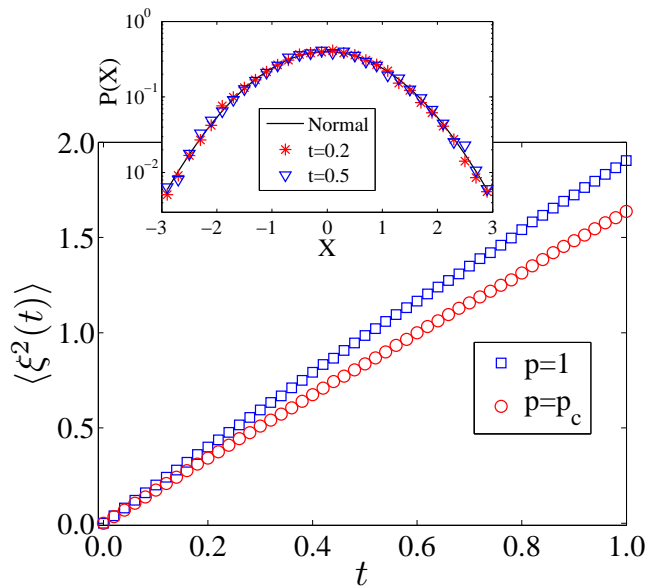} 
\caption{\label{fig::fig4} 
(color online) The statistics of the obtained driving function for LERW$_p$. 
\textbf{Main frame}: The second moment of the driving function $\langle \xi^2(t)\rangle$ versus Loewner time $t$ for normal LERW (i.e. $p=1$) and for LERW on critical percolation cluster. The obtained diffusion coefficients are $\kappa =1.68\pm 0.07$ and $\kappa=1.94\pm0.07$ for $p=p_c$ and $p=1$ respectively. \textbf{Inset}: Probability distribution of the driving function at two different Loewner times for LERWs on critical percolation cluster. The rescaled parameter $X$ is defined as $X=\xi(t)/\sqrt{\kappa t}$, where we have taken $\kappa=1.68$. The solid line is the normal distribution of vanishing mean value and unit dispersion. Results are averages over $4\times10^4$ realizations on a square lattice with $L=1025$ for both cases.} 
\end{figure}

Consider a random non-self-crossing curve SLE curve $\gamma(t)$, which starts at a point on the real axis and grows to infinity inside a region of the upper half plane $\mathbb{H}$. We parameterize the curve with the dimensionless parameter $t$, typically called Loewner time. At each time $t$, the $\mathbb{H}$ minus the curve $\gamma (t)$ can be mapped back to the $\mathbb{H}$ by a unique function $g_t(z)$, where $z$ is a point on $\mathbb{H}$ (its real and imaginary parts are denoted by $\Re z$ and $\Im z$, respectively). This function satisfies the Loewner equation \cite{Loewner23}, 
\begin{equation} 
\partial_{t}g_{t}(z)=\frac{2}{g_{t}(z)-\xi_{t}}, 
\end{equation} 
where the initial condition is $g_{t=0}(z)=z$ and $\xi_{t}$ is a continuous real valued function called the {\it driving function}. The driving function is proportional to the Brownian motion $B_t$ i.e. $\xi_t=\sqrt{\kappa} B_t$ if and only if the probability measure of $\gamma(t)$ satisfy conformal invariance and domain Markov property \cite{Schramm00, Schramm01}. This type of conformal curves is known as chordal SLE$_{\kappa}$. In addition to the chordal SLE, there is another type of SLE known as dipolar which joins the origin to a point of the line $\Im z = \pi$ (in the strip geometry) is described by a Loewner-type equation \cite{Bauer05}. 
As discussed before, the transformed LERW$_p$ are {\it dipolar} curves, which start at one point on the lower boundary and end when they touch a point on $y=L/2$, for the first time. We scale the LERW$_p$ curves by factor $2\pi/L$ to be in a strip $ 0< \Im z < \pi$. 
To more carefully inspect the correspondence with SLE, we use discrete version of Loewner equation for dipolar random curves to map the transformed LERW$_p$, represented by sequences of points $z_{i}$, $i=1,2,\ldots N$, onto a real-valued sequence $\xi(t_{i})$ defined at discrete $t_{i}$. 
Initially, we set $t=0$ and $\xi(0)=0$. We then apply a sequence of slit maps obtained by considering of a piecewise constant for the driving function at each step. At each iteration $i$, we map the point $z_i$ to the real axis at $\xi_{i}$ defined at discrete $t_i=t_{i-1}+\delta_i$ and we transform other points $z_j$ (for $j>i$) of the curve using the map appropriate for dipolar SLE \cite{Bauer05, Bernard07}, 
\begin{eqnarray} 
\delta t_{i}& = & 
- 2 ln [\cos(\frac{\Im z_i}{2})]\ \ ;\ \ 
\xi_{i} = \Re z_{i}\label{eq::slit.map}\\ 
z_{j} & = & 
\xi_{i}+ 
2\cosh^{-1}\left\{ 
\cosh\left[\frac{(z_{j} - \xi_i)}{2}\right] 
/\exp(-\delta t_{i}/2)\right\}.\nonumber 
\end{eqnarray} 
This map converges to the exact one for vanishing $\delta t$ \cite{Bauer03}. Here, we restrict our attention to the LERW on critical percolation cluster. For comparison, we also study LERW on Euclidean lattice (i.e. $p=1$). We took $4 \times 10^4$ disorder realizations of LERW in a lattice with $L=1025$ for $p=1$ and $p=p_c$. The average over realizations of the disorder of $\langle \xi^2(t)\rangle$ versus Loewner time $t$ in dipolar LERW is plotted. The obtained diffusion coefficients are $\kappa =1.68\pm 0.07$ and $\kappa=1.94\pm0.07$ for $p=p_c$ and $p=1$ respectively. To confirm the Gaussianity of the driving function $\xi_t$, the probability distribution for the rescaled driving function \mbox{$X=\xi(t)/\sqrt{\kappa t}$} for two different times for the LERW is plotted in the inset of Fig.~\ref{fig::fig4}. This results indicate that statistics of $\xi(t)$ converges to a Gaussian process with zero mean and $1.68\pm 0.07$ variance, in good agreement with the results discussed above. We also study the correlation function $C(n)=\langle[\xi(t_{i+n+1})-\xi(t_{i+n})][\xi(t_{i+1})-\xi(t_{i})]\rangle$ at intermediate times to test the Markovian property for $\xi(t)$; it decays rapidly for both cases. It is also important to remark that the slit map goes to real mapping only in the continuum limit and converges for sufficiently small $\delta t_i$ \cite{Bauer03}. Due to these strong discretization effects, the numerical results obtained with the direct SLE method are less precise than the other two methods (winding angle and left-passage), as is well known in the literature \cite{Daryaei12, Bernard07, Bernard06}. For both of fractal ($p=p_c$) and Euclidean geometries ($p>p_c$), within the error bars, the results we have obtained for $\kappa$ are in agreement with the ones obtained with the fractal dimension, winding angle, and left-passage probability. 
\section{Summary and Discussion} 
In this paper we mainly study the scaling limit of LERW$_p$ on percolation cluster, with occupation probability above and equal to the critical value, $p\geq p_c$. The SLE tests indicate that scaling limit of this model for $p>p_c$ is SLE$_2$. Although our study in this regime is restricted to a few points, it has recently been shown in Ref. \cite{Yadin11} that if the scaling limit of the RW on a planar graph is planar Brownian motion, such as percolation cluster for $p>p_c$, then the scaling limit of its loop-erasure is SLE$_2$ which confirms our results in this regime. However LERWs on critical percolation are likely SLE curves with \mbox{$\kappa=1.732\pm0.016$}. This value, similar to our recent finding in watershed model \cite{Daryaei12}, is outside of the well-known duality conjecture range $2\leq\kappa\leq8$. 
Near the percolation threshold, $p_c$, there is a crossover regime, shown in Fig.\ref{fig::fig1}, from {\it Euclidean} to {\it fractal} geometry. To achieve a better understanding of this regime, we have also investigated how the winding angle of the LERW$_p$ crosses over between these two universality classes by gradually decreasing the value of the parameter $p$ from 1 to $p_c$. Our findings for crossover regime, shown in Fig.\ref{fig::fig2}, clearly demonstrates that for finite systems, two crossover exponents and a scaling relation can be derived. 
A famous relation between central charge of conformal models which possess a second level null vector in their Verma module and the diffusivity $\kappa$ is $c=(3\kappa-8)(6-\kappa)/2\kappa$ \cite{Bauer03}. If the LERW on critical percolation cluster is conformally invariant it likely corresponds to a logarithmic CFT (LCFT) with central charge $c=-3.45\pm0.10$. In particular, LERW on Euclidean lattice is believed to have $c=-2$. It is also noteworthy that negative central charges have been reported in different contexts like, e.g., stochastic growth models, $2D$ turbulence, and quantum gravity \cite{Duplantier92, *Flohr96, *Lipatov99}. However, the conformally invariant of the LERW$_p$ on percolation cluster cannot be comprehended as a strong proof. Nevertheless, if such invariance is established, it becomes possible to develop a field theory for this new universality class. Moreover, due to the connection between LERW and other important statistical models, and also some mathematical constructions for this model, it is possible to find exact results regarding the existence of conformal invariance and scaling properties. 
In addition to the conformal symmetry, it is needed that the LERW$_p$ curves possess a domain Markov property in the scaling limit to be SLE. However, the direct examination of domain Markov property numerically is an extremely challenging task, and only a few numerical studies have been tested it non-rigorously \cite{Bernard07}. 
Here, we did not attempt to check the domain Markov property of the LERW$_p$ curves. Instead, we simply tested the Markovian property of $\xi(t)$; since at each time $t$, there is a unique conformal map which takes the LERW curve to a real function $\xi(t)$ on the real axis, it is expected that the Markovian property of $\xi(t)$ is as a result of the domain Markov property of LERW$_p$. 
The connection between SLE and statistical properties of LERW$_p$ provides a new perspective to look at such random path with a new eye and to build bridges between connectivity in disordered media and other research areas in mathematics, percolation, and quantum field theory. This work opens up several challenges. Besides the need to examine directly both of conformal invariance and the domain Markov property, it would be interesting to formulate a CFT scheme in a fractal geometry. Finally, scaling limit of obtained LERW from an anomalous diffusion on fractal landscape is still an important open question. 
\begin{acknowledgments} 
We are grateful to S. Rouhani and S.Moghimi-Araghi for helpful comments on the manuscript. 
\end{acknowledgments} 

\bibliography{LERW} 
\end{document}